\documentclass[final,authoryear,1p,times]{elsarticle}

\usepackage{amsmath,amssymb,xspace,subfigure}
\usepackage{amsfonts}
\usepackage[dvipsnames]{xcolor}
\usepackage[]{graphicx}
\usepackage{url}
\PassOptionsToPackage{
  colorlinks=true,
  citecolor=blue,
  linkcolor=blue,
  urlcolor=teal
}{hyperref}
\usepackage{natbib}
\usepackage{hyperref}
\usepackage{fullpage}
\usepackage{cases}
\usepackage{setspace}
\newcommand{\epsth}{$\epsilon_0$}

\graphicspath{{./Figures/}}

\journal{Elsevier}

\begin{document}

\begin{frontmatter}



\title{Midveins regulate the shape formation of drying leaves}


\author[a]{Kexin Guo\fnref{label2}}
\author[b]{Yafei Zhang\fnref{label2}}
\author[a]{Massimo Paradiso}
\author[c,d]{Yuchen Long}
\author[a,e]{K. Jimmy Hsia\corref{cor5}}
\author[f]{Mingchao Liu\corref{cor5}}
\cortext[cor5]{Corresponding authors: kjhsia@ntu.edu.sg (K.J.H.), m.liu.2@bham.ac.uk (M.L.)}
\fntext[label2]{K.G. and Y.Z. contributed equally to this work.}

\address[a]{School of Mechanical and Aerospace Engineering, Nanyang Technological University, Singapore 639798, Singapore}
\address[b]{Racah Institute of Physics, The Hebrew University of Jerusalem, Jerusalem 91904, Israel}
\address[c]{Department of Biological Sciences, National University of Singapore, Singapore 117543, Singapore}
\address[d]{Mechanobiology Institute, National University of Singapore, Singapore 117411, Singapore}
\address[e]{School of Chemistry, Chemical Engineering and Biotechnology, Nanyang Technological University, Singapore 637371, Singapore}
\address[f]{Department of Mechanical Engineering, University of Birmingham, Birmingham B15 2TT, UK}

\begin{abstract}

Dried leaves in nature often exhibit curled and crumpled morphologies, typically attributed to internal strain gradients that produce dome-like shapes. However, the origin of these strain gradients remains poorly understood. Although leaf veins—particularly the midvein—have been suggested to influence shape formation, their mechanical role has not been systematically investigated. Here, we demonstrate that mechanical constraints imposed by the midvein play a crucial role in generating the diverse morphologies that emerge during leaf drying. Combining numerical simulations and theoretical analysis, we show that a uniformly shrinking leaf lamina constrained by a non-shrinking midvein gives rise to two distinct types of configurations: \textit{curling-dominated} and \textit{folding-dominated} morphologies. In the \textit{curling-dominated} regime, both \textit{S}-curled and \textit{C}-curled shapes emerge, with \textit{C}-curled configurations more commonly observed due to their lower elastic energy. In contrast, the \textit{folding-dominated} regime features folding accompanied by edge waviness. Theoretical modeling reveals a linear relationship between midvein curvature and mismatch strain, consistent with simulation results. Moreover, we find that the morphological outcome is governed by the ratio of bending stiffnesses between the lamina and the midvein. We construct a comprehensive phase diagram for the transitions between different configurations. These findings provide a mechanical framework for understanding shape formation in drying leaves, offering new insights into natural morphing processes and informing the design of bio-inspired morphable structures.

\end{abstract}


\end{frontmatter}

\section{Introduction} \label{sec:intro}

\begin{figure}[bth]
    \centering
    \includegraphics[width=1.0\textwidth]{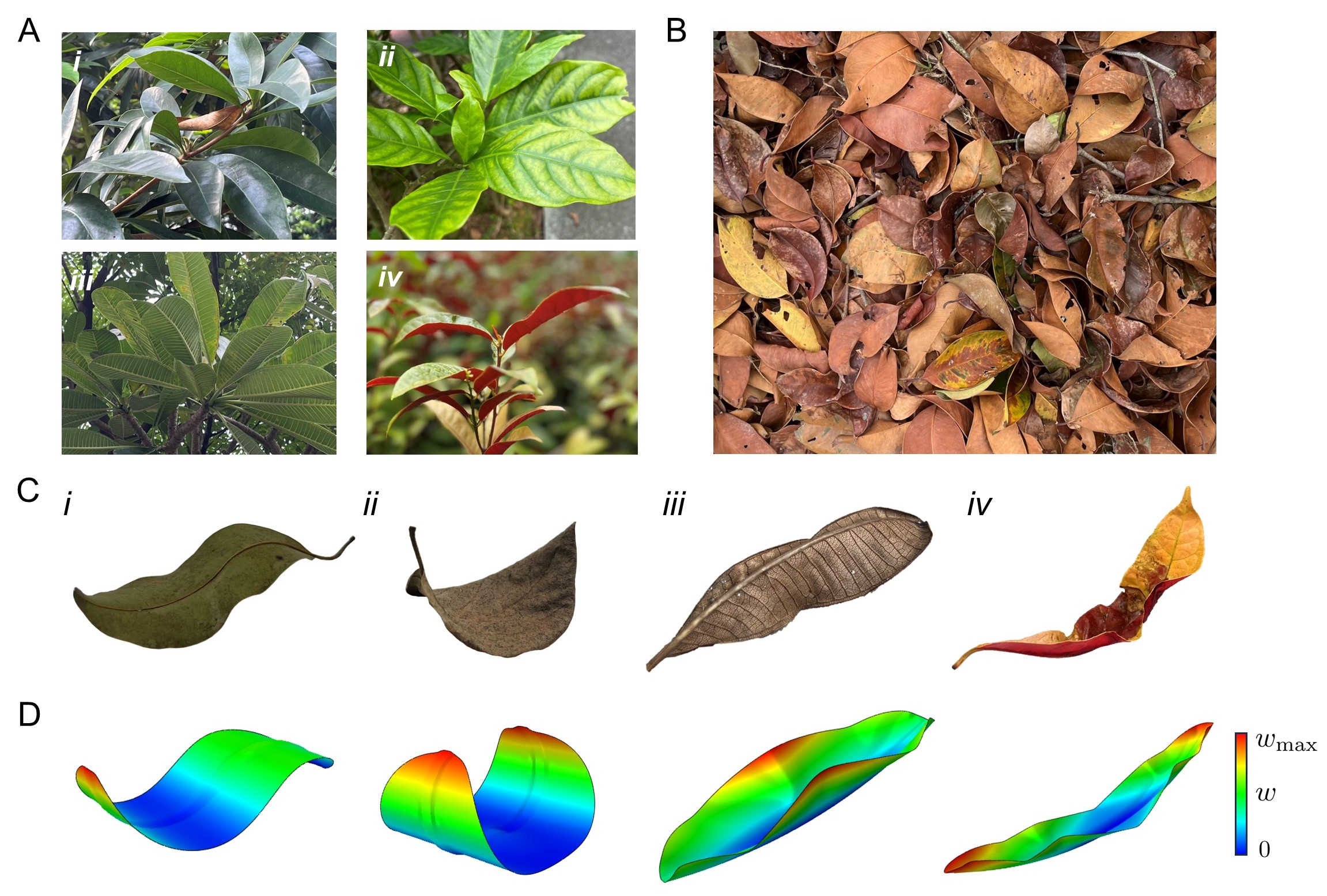}    
    \caption{\textbf{Drying leaves display diverse shape transformations influenced by midvein constraints.} 
    \textbf{A.} Representative plant species and their fresh leaves: \textit{(i)} \textit{Gardenia jasminoides}, \textit{(ii)} \textit{Cyrtophyllum fragrans}, \textit{(iii)} \textit{Plumeria obtusa}, and \textit{(iv)} \textit{Excoecaria cochinchinensis}. 
    \textbf{B.} Fallen leaves collected from the ground display a variety of morphologies, including curling, folding, and combined folding with edge waviness. 
    \textbf{C.} Typical dried leaves from the corresponding species shown in \textbf{A}, labeled \textit{i}–\textit{iv}. 
    \textbf{D.} Simulation results of drying leaf morphologies with elliptical laminae constrained by a thickened midvein, based on the model proposed in this study detailed in later sections. The four configurations correspond to the respective real leaf morphologies for the species in \textbf{C}. 
    Color maps indicate out-of-plane displacement in the $z$-direction \(w\), with red denoting the maximum magnitude.}    
    \label{fig:overview}
\end{figure}

The rich diversity of plant leaf morphologies, widely observed in nature (Fig.~\ref{fig:overview}A), has long captivated researchers across disciplines including biology \citep{du2018-molecular}, physics \citep{lewicka2011-vonKarman}, and engineering \citep{yang2023-morphing}. Recent studies have extensively investigated the key factors regulating leaf morphogenesis, aiming to uncover the interplay between growth and form. Beyond identifying crucial genes and molecular regulators, it has become clear that mechanical factors also provide mechanisms for translating molecular growth processes into macroscopic shapes \citep{guo2022}. Various three-dimensional (3D) leaf forms have been attributed to buckling of the lamina induced by differential growth strains—for example, the saddled shapes and edge waviness in long leaves caused by excess edge growth \citep{maha2009-longleaf,maha2011-lily,huang2018-differential}, or the curved surfaces of lotus leaves shaped by stem constraints and interfacial forces from the supporting water substrate \citep{xu2020-lotus}. These studies highlight the critical role of mechanical deformation and instability in generating the rich morphological diversity of plant leaves. In addition to intrinsic growth patterns, internal physical constraints, such as the midvein and stem, further modulate the final equilibrium shape of the lamina \citep{xu2020-lotus,wang2024-morpho}, although their precise roles remain incompletely understood.

Besides the fascinating behaviors observed in leaf morphogenesis, aesthetically intriguing fallen leaves also exhibit rich morphing phenomena (Fig.~\ref{fig:overview}B). During senescence, leaves undergo marked changes in both color and shape, typically becoming curled, folded, or crumpled upon drying. Examples from four representative species are shown in Fig.~\ref{fig:overview}C, highlighting two types of morphologies -- curling-(i,ii) and folding-(iii,iv) dominated. Based on a model proposed in this paper, these morphologies are also successfully reproduced through simulations (Fig.~\ref{fig:overview}D). This raises a natural question: how do initially flat leaves transform into such diverse, often non-flat shapes upon drying? While senescence involves complex biochemical and physiological processes \citep{woo2019-senescence}, water loss—often exceeding 50\% of the initial leaf mass \citep{tomaszewski2016}—plays a central  role, as it leads to lamina shrinkage giving rise to mechanical deformation \citep{lewicki1998, guo2024dehydration}. These post-mortem deformations offer a natural model for investigating how internal stresses, material heterogeneity, and geometric constraints drive complex shape transformations in plant leaves \citep{Motala2015,Carvajal2024}.
Understanding the mechanisms that regulate these morphing modes not only deepens our knowledge of mechanical instabilities in biological organs, but also guides the design of soft materials and biomimetic structures, as well as strategies for leaf-inspired adaptation.

Early attempts to simulate drying leaf shapes considered constrained deswelling of gel films \citep{liu2010-modeling}, followed by studies on drying-induced deformations under differential tissue contraction, where a linear strain gradient was applied to examine the effects of aspect ratio and normalized size on dome-shaped curvature \citep{xiao2011}. These studies also demonstrated the influence of venation patterns through simulations based on real leaves. In the computer graphics community, \citet{jeong2013} developed a biologically inspired approach that models the full leaf surface as a double-layered mesh aligned with vein structures, capturing sharp creases and complex curling by incorporating osmotic water flow and heterogeneous shrinkage. Although these studies shed light on the roles of venation and water transport, they fall short of offering a mechanistic explanation for the diverse morphologies observed in drying leaves. In nature, several typical morphologies recur, notably folding along the midvein—a pattern reminiscent of behaviors in locally stretched elastic sheets \citep{guo2025localized}. This similarity prompts the question of whether comparable mechanisms, driven by local constraints imposed by the midvein, also govern shape transformations in natural leaves.

Leaf veins impose strong physical constraints to the lamina during shape morphing. Among the hierarchical vascular structures, the first-order vein (also called the midvein or midrib) is a widely-preserved primary vascular structure that provides pathways for water transport and mechanical support to the leaf blade, but also  as a physical constraint to leaf deformations. It runs from the base of the leaf to the apex \citep{sack2013-venation}, and is significantly thicker than the secondary veins that distribute relatively uniformly in the lamina. Moreover, the vascular tissues of midveins have higher elastic moduli than the mesophylls due to the dense cellulose microfibrils and spring-like microstructure that resists deformation along the longitudinal direction \citep{gibson2012hierarchical}. Therefore, a leaf blade can be regarded as a composite structure of an isotropic plate connected to a beam-like midvein. Studies on leaf shape formation mainly regard the leaf as an isotropic uniform thin sheet \citep{maha2009-longleaf,maha2011-lily}, ignoring the midvein's effect. In fact, leaf midveins may have significant impacts on leaf morphogenesis, since leaf laminae and veins are tightly coupled at their interface. \cite{wang2024-morpho} discussed how constraints from the midvein affect the morphogenesis of curled leaves. The interaction between leaf lamina and midvein is not yet fully understood in the context of shape morphing, especially during the drying process.

In this work, we investigate the mechanisms and key factors governing the shape formation of drying leaves constrained by a midvein. We propose a general mechanical framework in which the midvein acts as a structural constraint on the shrinking leaf lamina, leading to differential-strain-induced buckling and the emergence of characteristic morphologies. Using a non-Euclidean elasticity model, we successfully capture the configurations observed in  drying leaves. Section~\ref{sec:system} presents the system setup, along with the mathematical framework and numerical simulation methods. Based on these formulations, Section~\ref{sec:results} analyzes the underlying physical mechanisms, derives scaling laws predicting the midvein curvature and the critical buckling strain, and further validates the theoretical predictions through finite element method (FEM) simulations. We also explore the influence of key parameters—the bending rigidity ratio between the midvein and lamina, and their thickness ratio—on the resulting morphologies, culminating in a phase diagram of morphological transitions as a function of system parameters. Together, these results highlight the critical mechanical role of leaf midveins in shape formation and offer new insights into the physical processes underlying leaf drying and senescence. Finally, Sections~\ref{sec:discussion} and \ref{sec:conclusion} provides additional discussions and concluding remarks.

\section{The Model}\label{sec:system}

\begin{figure}[ht]
    \centering
    \includegraphics[width=.9\textwidth]{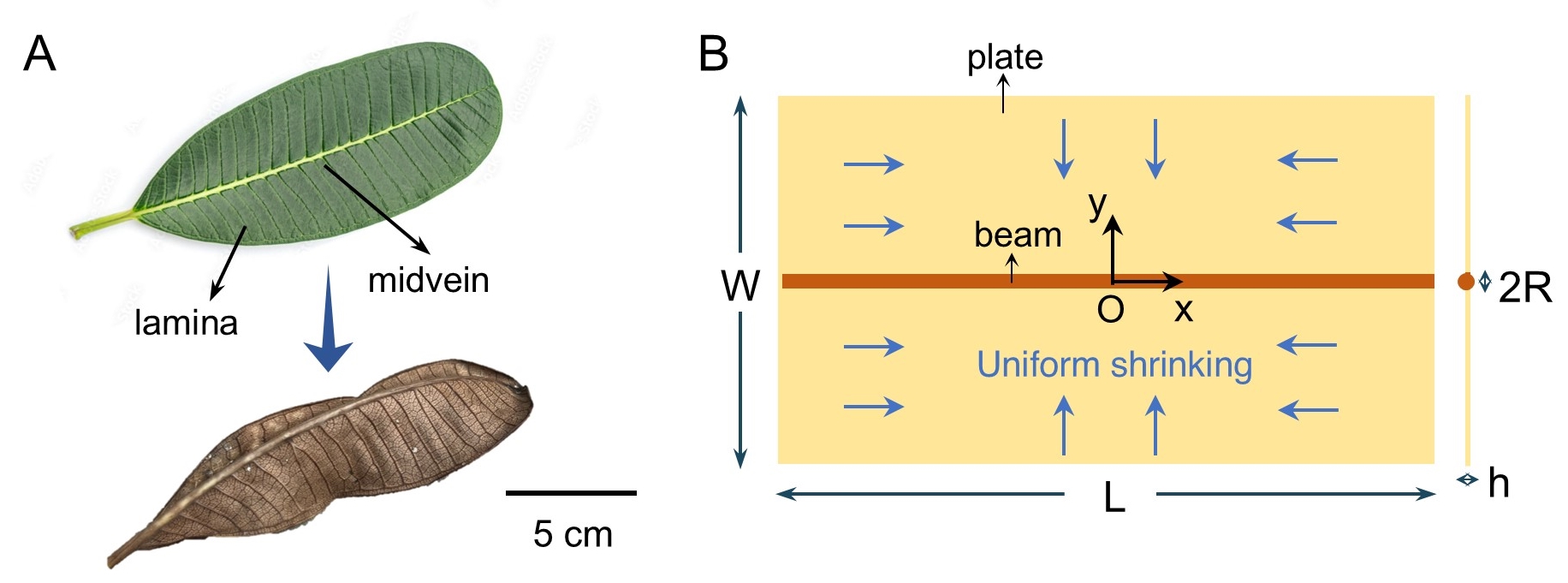}
    \caption{\textbf{Simplified model capturing shape transformation in drying leaves.} 
    \textbf{A.} A frangipani (\textit{Plumeria obtusa}) leaf that is flat in the fresh state (top) transforms into a folded and wavy 3D shape when dried (bottom), with a visible structural distinction between the lamina and midvein. 
    \textbf{B.} Schematic of a simplified leaf structure: the lamina is modeled as a thin plate (yellow) with width $W$, length $L$, and thickness $h$, embedded with a central midvein represented as a cylindrical beam (brown) of radius $R$. Upon drying, the lamina undergoes uniform isotropic shrinking, while the midvein remains undeformed.}
    \label{fig:model}
\end{figure}

In this study, we approximate the complex structure of a natural leaf (Fig.~\ref{fig:model}A) as a simplified composite model that captures the essential mechanisms driving its shape formation and evolution during the drying process. As illustrated in Fig.~\ref{fig:model}B, the leaf is represented by two primary components: the lamina and the midvein. The lamina is modeled as a thin plate with width $W$, length $L$, and thickness $h$, undergoing isotropic, uniform shrinkage as it loses water; whereas the midvein is treated as a cylindrical beam of radius $R$ that resists contraction, acting as a structural constraint. This differential response between the lamina and midvein induces strain mismatch, generating internal stresses and buckling instabilities that ultimately drive the formation of the diverse 3D morphologies observed in drying leaves (Fig.~\ref{fig:overview}). Our theoretical analysis, based on the non-Euclidean elasticity framework~\citep{efrati2009elastic}, together with  FEM simulations, provides a quantitative description of how geometric and material parameters govern the overall shape transformation, showing good agreement with experimental observations (Fig.~\ref{fig:overview}C, D).

\subsection{Theoretical framework}\label{sec-theory}
Non-Euclidean elasticity provides a natural framework for modeling thin solids with incompatible intrinsic geometries—such as those arising from differential growth, swelling, or shrinkage—where the prescribed distances and angles cannot be embedded in Euclidean space without distortion \citep{efrati2009elastic}. This geometric incompatibility is encoded in a reference metric tensor $\bar{a}$ and a reference curvature tensor $\bar{b}$, which together define a target configuration that may not admit a smooth isometric immersion into three-dimensional space \citep{sharon2010mechanics}. As a result, residual stresses arise even in the absence of external loads, driving the system to deform out of plane \citep{sharon2007,xu2020-lotus,maha2009-longleaf,maha2011-lily} or form localized structures \citep{zhang2025geometrically}. In drying leaves, the mismatch between the shrinking lamina and the relatively undeformed midvein gives rise to exactly this type of intrinsic incompatibility.

We formulate the drying leaf as an non-Euclidean plate, in which the leaf is modelled as a rectangular plate of width $W$, length $L$ and thickness $h$, representing the lamina, with an embedded central region of width $2R$ representing the midvein (Fig.~\ref{fig:model}B). Specifically, this two-dimensional domain in Fig.~\ref{fig:model}B is prescribed with a reference first fundamental form, $\bar{a}$, and a reference second fundamental form, $\bar{b}$, as the target (stress-free) configuration,  
\begin{equation}
\label{eq:refGeo}
    \bar{a} = \eta^2(y)\begin{bmatrix}1 &0 \\0 & 1\end{bmatrix}, \quad \bar{b} = 0.
\end{equation}
In the framework of non-Euclidean elasticity, the first and second fundamental forms are symmetric second-order tensors. The first fundamental form encodes the intrinsic geometry of the leaf surface by defining the lengths of infinitesimal line elements in the tangent plane, while the second fundamental form describes the extrinsic geometry by characterizing the variation of the surface normals along the leaf.

Following the drying process, the lamina region (where $R<|y|<W/2$) contracts by a factor of $1-\eta_0$, while the midvein retains its original length. Consequently, the shrinkage function $\eta(y)$ in the reference metric $\bar{a}$ can be expressed using the Heaviside step function $H(x)$ as
\begin{equation}
\label{eq:shrinkage}
    \eta(y) = \eta_0+(1-\eta_0)H(R-|y|).
\end{equation}

After applying the shrinkage function, the reference domain $\mathcal{D} = [-L/2,L/2]\times[-W/2, W/2]$ (Fig.~\ref{fig:model}B) deforms into an actual configuration characterized by the metric $a$ and curvature $b$. Assuming a quasi-static shrinkage process and neglecting changes in Young's modulus $E$ and Poisson's ratio $\nu$ during drying, the elastic energy consists of two contributions: the stretching energy density, $W_s \sim E h \lvert a-\bar{a}\rvert ^2$, penalizes deviations of the actual metric from the reference metric; while the bending energy density, $W_b \sim E h^3 \lvert b-\bar{b} \rvert^2$, penalizes the curvature mismatch. Consequently, the total elastic energy of the system is given by~\citep{efrati2009elastic},
\begin{equation}
\label{eq:totalEnergy0}
    \mathcal{U} = \frac{E}{8(1-\nu^2)} \int_{\mathcal{D}} \left(h \lvert a-\bar{a}\rvert^2 +\frac{h^3}{3}\lvert b-\bar{b}\rvert^2 \right) dS,
\end{equation}
where $dS$ is the area element, and the notation $|T|$ represents the elastic norm of a tensor $T$, with $|T|^2 \triangleq \nu \operatorname{Tr}^2\left[\bar{a}^{-1} T\right]+(1-\nu) \operatorname{Tr}\left[\bar{a}^{-1} T\right]^2$ and $\operatorname{Tr}\left[\cdot\right]$ being the trace operator.

The reference (or target) configuration, represented by $\bar{a}$ and $\bar{b}$, can be independently assigned based on the physical process. In contrast, the actual equilibrium configuration, described by $a$ and $b$, must satisfy the geometric compatibility conditions inherent to non-Euclidean surfaces \citep{efrati2009elastic}. Consequently, the equilibrium state is determined by minimizing the total energy $\mathcal{U}$ with respect to $a$ and $b$. In practice, this complex minimization can be typically carried out using numerical methods, such as the FEM analysis, where the reference geometry is prescribed by the material's thermal properties. Specifically, the applied thermal strain (shrinkage strain) is set as 
\begin{equation}
\label{eq:thermalstrain}
    \epsilon(y) = \epsilon_0-\epsilon_0H(R-|y|).
\end{equation}
with $\epsilon_0 \triangleq 1-\eta_0$ in Section~\ref{sec:results} (e.g., Figs.~\ref{fig:fem}B,E).

\subsection{Finite Element Method (FEM) Simulation}

\begin{figure}[ht]
    \centering    
    \includegraphics[width=1.0\linewidth]{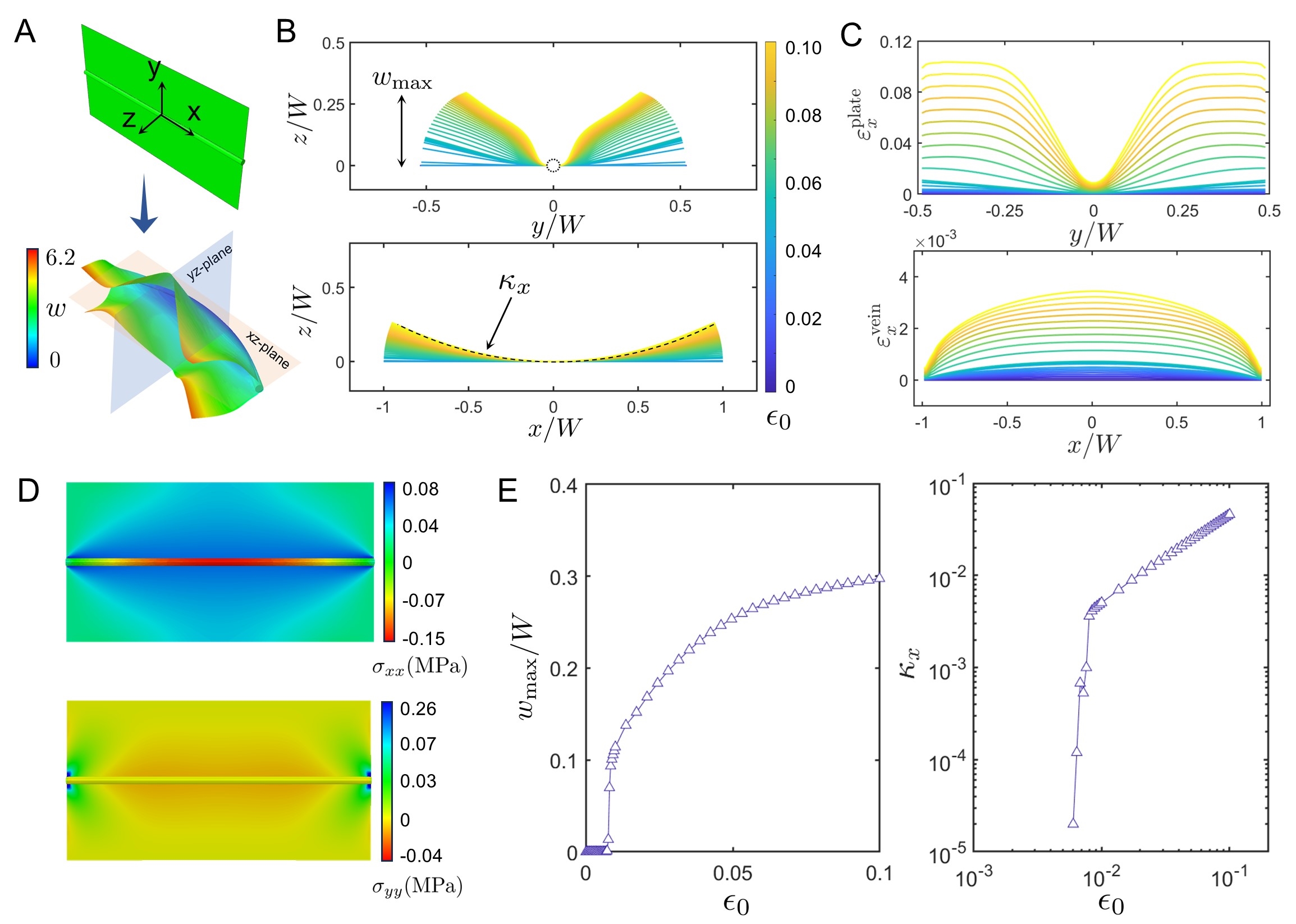}
    \caption{\textbf{FEM simulation of a simplified drying leaf structure.} 
    \textbf{A.} FEM model consisting of a central beam (midvein) embedded in a thin rectangular plate (lamina), where the lamina is subjected to a prescribed shrinkage strain \(\epsilon_0 = 0.1\). Contour colors represent the out-of-plane displacement in the $z$-direction \(w\). 
    \textbf{B.} Deformed profiles of the lamina (top; cross-section along the \(yz\)-plane) and the midvein (bottom; cross-section along the \(xz\)-plane) corresponding to \textbf{A}. The color bar indicates varying magnitudes of shrinkage strain \(\epsilon_0\). Key geometric measures \(w_{\mathrm{max}}\) and \(\kappa_x\) are labeled. 
    \textbf{C.} Distributions of the shrinkage strain in the \(x\)-direction: in the lamina (\(\epsilon_x^{\text{plate}}\), top) and in the midvein (\(\epsilon_x^{\text{vein}}\), bottom). 
    \textbf{D.} Stress components: longitudinal stress \(\sigma_{xx}\) (top) and transverse stress \(\sigma_{yy}\) (bottom). Negative values indicate compressive stress. 
    \textbf{E.} Quantification of the deformation response as a function of shrinkage strain \(\epsilon_0\): normalized maximum deflection \(w_{\mathrm{max}}/W\) (left), and midvein curvature \(\kappa_x\) (right). The model geometry is defined by \(W = 10\), \(L = 20\), \(h = 0.075\), and \(R = 0.25\).}
    \label{fig:fem}
\end{figure}

To obtain solutions of the equilibrium configurations developed in the preceding analysis, we use FEM simulation to model the leaf structure numerically and to minimize Eq.~(\ref{eq:totalEnergy0}) in the general cases. FEM simulation is conducted using the commercial software ABAQUS/2023. The model consists of a rectangular plate with width $W$, length $L$, and thickness $h$, embedded with a solid cylindrical beam of radius $R$ as illustrated in Fig.~\ref{fig:fem}A (top). The plate is modelled using quadrilateral plane-stress elements (S4R), and the solid beam is modelled using solid elements (C3D8R). Nodes on the edges of the plate are tied to the edges of the midvein by fixing all degrees of freedom. The entire structure is free-standing by limiting only the rigid body displacement. Linear elastic constitutive relations are applied: Poisson’s ratio $\nu=0.48$ is used since the plant tissues are considered incompressible; Young's modulus is set to $E_v=1~\rm{GPa}$ for the midvein (solid beam) and $E_l=30~\rm{MPa}$ for the lamina (plate) for most of the simulations unless otherwise stated. Shrinking is imposed to only the laminae by a uniform isotropic contracting thermal strain, \epsth. In the current model, buckling can be induced without applying geometric imperfections, and thus artificial imperfections are not introduced for all simulations.

\section{Results and analyses}\label{sec:results}

\subsection{Shape morphing by dyring-induced strain-mismatch}

In this section, we first demonstrate the results of the FEM simulation of equilibrium configurations after constrained leaf shrinking. As mentioned in Section~\ref{sec-theory}, a contracting thermal strain \epsth~(see Eq.~(\ref{eq:thermalstrain})) is applied uniformly to the lamina. Figure~\ref{fig:fem} shows the simulation result for a model in which the midvein is significantly more rigid than the lamina (see figure captions for more details). The equilibrium configuration shown in Fig.~\ref{fig:fem}A is characterized by bending of the midvein as well as folding of the lamina. The projection of deformed shape on the \textit{xz}-plane and that on  \textit{yz}-plane  are illustrated with increasing strains in Fig.~\ref{fig:fem}B. The corresponding strain distributions are shown in Fig.~\ref{fig:fem}C. 
The strain applied here is the reference state, exhibited in the lamina when it is not constrained by the midvein. This can be seen from Fig.~\ref{fig:fem}C (top) --- the equilibrium or actual strains in the lamina converge to its reference state as it is less constrained towards the edges. Upon drying, the midvein undergoes little deformation, constraining the lamina in its proximity. This strain mismatch leads to the formation of a transition layer \citep{chen2022formation}, within which the strain changes from $0$ to \epsth. 
Meanwhile, the midvein is under axial compression from the lamina connected, as shown in Fig.~\ref{fig:fem}C (bottom). The middle portion of the midvein bears the largest compressive strain. This is consistent as the strain state of a fibre bound within a soft matrix under compression \citep{zhao2016buckling}. The simulations show that the stress along the longitudinal direction in the lamina, $\sigma_{xx}$, is tensile, while the transverse stress, $\sigma_{yy}$, is compressive at some distances away from the short ends (Fig.~\ref{fig:fem}D). This compressive stress therefore led to the onset of out-of-plane buckling for lamina with small thickness at some critical strain -- to relieve the internal stresses, which otherwise would result in tremendous strain energy in the lamina (the details will be  discussed in Section \ref{sec:theory_shape_formation}). The onset of buckling at a critical strain $\epsilon_0^c$ is observed from the sudden increase in out-of-plane deflection $w_{max}$ of the lamina (Fig.~\ref{fig:fem}E, left), or the increase in midvein curvature $\kappa_x$ (Fig.~\ref{fig:fem}E, right). Meanwhile, we also observe the midvein curvature exhibits a power-law growth with respect to the shrinkage strain, $\epsilon_0$, which will also be analysed in Section \ref{sec:theory_shape_formation}.

\subsection{Shape evolution with increasing shrinkage strain}

\begin{figure}[t]
    \centering
    \includegraphics[width=1.0\textwidth]{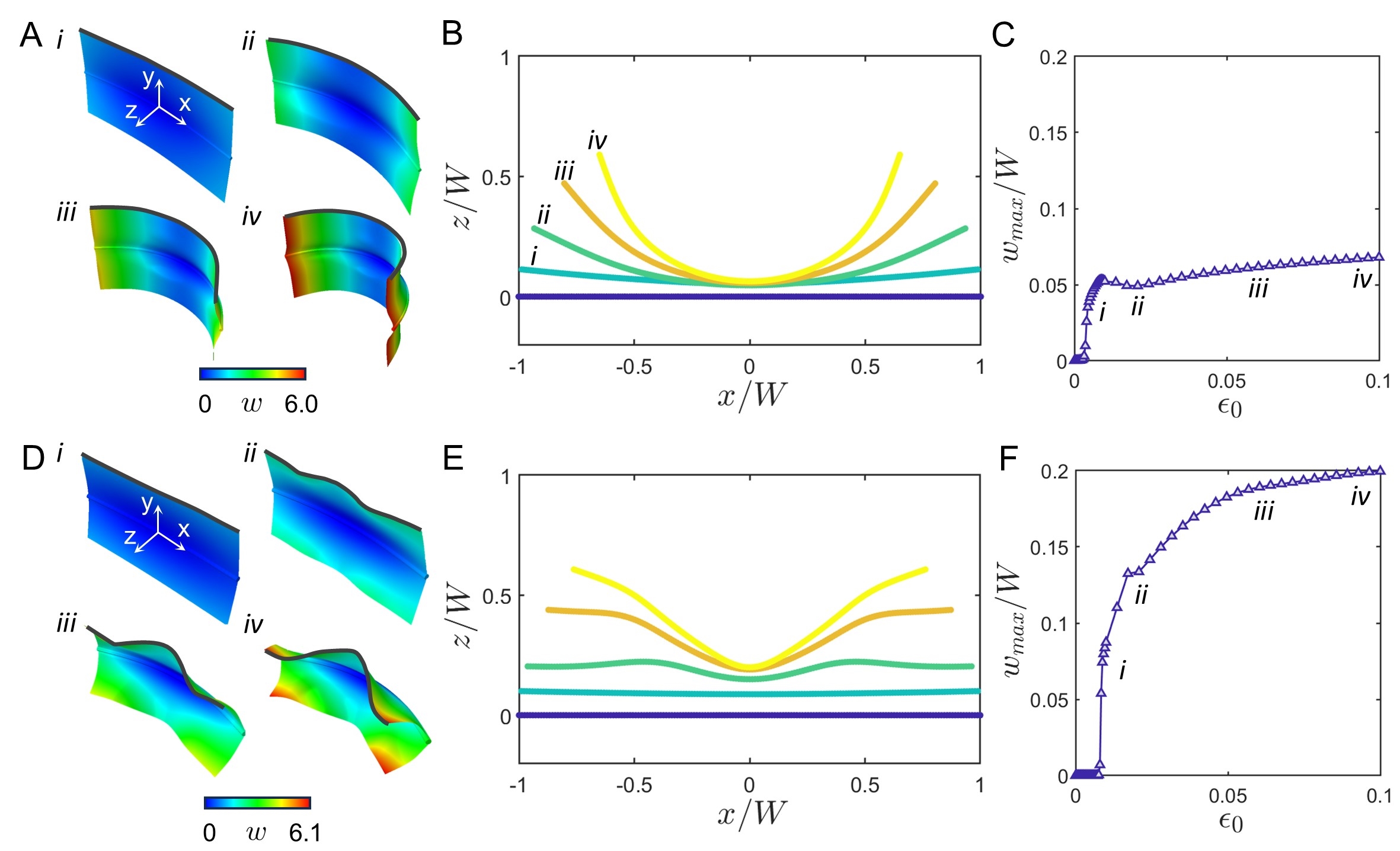}
    \caption{\textbf{Evolution of morphology with increasing shrinkage strain in the lamina.} 
    \textbf{A–C.} Simulations of leaf curling with a smaller midvein-to-lamina thickness ratio (\(R/h = 0.5\)).  
    \textbf{D–F.} Simulations of leaf folding with a larger thickness ratio (\(R/h = 2.5\)). In both cases, the plate dimensions are fixed as \(L/W = 2\), \(h/W = 1\times10^{-3}\).  
    \textbf{A, D.} Deformed configurations at shrinkage strains, \(\epsilon_0\): (i) 1\%, (ii) 2\%, (iii) 6\%, and (iv) 10\%. Color contours represent the out-of-plane displacement \(w\).  
    \textbf{B, E.} Edge profiles extracted at \(y = W/2\), corresponding to configurations (i–iv).  
    \textbf{C, F.} Normalized maximum deflection of the \(yz\)-cross section, \(w_{\mathrm{max}}/W\), plotted as a function of the shrinkage strain \(\epsilon_0\).}
    \label{fig:evolve}
\end{figure}

Here we examine the evolution of two primary morphological modes—\textit{curling-dominated} and \textit{folding-} \textit{dominated}—as observed in drying leaves (see Fig.~\ref{fig:overview}C), in response to increasing shrinkage strain. 
We begin with a leaf configuration featuring a significantly soft midvein (\(B_v/B_l = 0.02\)), modeled by setting the midvein-to-lamina thickness ratio to \(R/h = 0.5\), as illustrated in Fig.~\ref{fig:evolve}A. As the shrinkage strain in the lamina increases, the initially flat leaf gradually transitions into a curled shape, evolving from configuration (i) to (iv). To quantify this curling behavior, we extract the edge profile at \(y = W/2\), as shown in Fig.~\ref{fig:evolve}B. The blue straight line (\(z/W = 0\)) corresponds to the undeformed, flat configuration and serves as a reference. The plot of normalized maximum out-of-plane deflection of the \(yz\)-cross section, \(w_{\mathrm{max}}/W\), versus shrinkage strain \(\epsilon_0\) reveals a clear buckling transition from the flat to the curled state (Fig.~\ref{fig:evolve}C), with all configurations (i–iv) lying in the post-buckling regime.
In contrast, Fig.~\ref{fig:evolve}D shows results for a leaf with a relatively stiffer midvein (\(B_v/B_l = 12.89\)), corresponding to \(R/h = 2.5\). As the shrinkage strain increases from (i) to (iv), the shape evolves from flat to folded, and eventually develops edge waviness. This progression is evident in the edge profiles shown in Fig.~\ref{fig:evolve}E. Compared to the curling case in Fig.~\ref{fig:evolve}B, the folded configurations exhibit localized bending and sharp corners, especially pronounced at higher strain levels. The corresponding deflection plot in Fig.~\ref{fig:evolve}F again indicates a buckling transition, with all configurations (i–iv) lying beyond the critical strain in the post-buckling regime.
Building on these simulation results, we now proceed to develop a theoretical analysis to uncover the underlying mechanism of shape formation.

\subsection{ Theoretical Analysis}
\label{sec:theory_shape_formation}

\subsubsection{Geometric Origin of Shape Formation} \label{sec:theory-incompatibility}

To well-define a smooth surface in the configuration space (\textit{i.e.}, the 3D Euclidean space), three geometric compatibility conditions must be satisfied: one from Gauss's Theorema Egregium and two from the Mainardi–Codazzi equations~\citep{do2016differential}. Failure to meet any of these conditions results in geometric frustration and consequent mechanical instabilities~\citep{siefert2021euclidean}. In the reference configuration, Gauss's Theorema Egregium requires that the Gaussian curvatures, $K$, computed from the reference metric $\bar{a}$ and the reference curvature $\bar{b}$ coincide, i.e., $K({\bar{a}}) = K({\bar{b}})$. Meanwhile, the Mainardi–Codazzi equations imply that the curvature tensor is conservative, $\nabla_{\alpha}\bar{b}_{\beta\gamma}=\nabla_{\beta}\bar{b}_{\alpha\gamma}$.
Combining Eqs.~(\ref{eq:refGeo}) and (\ref{eq:shrinkage}), we obtain
\begin{equation}
    K(\bar{a}) = \frac{1}{2}\bar{a}^{\alpha\beta}\left(\partial_{\gamma}\Gamma_{\alpha \beta}^{\gamma} - \partial_{\beta}\Gamma_{\gamma \alpha}^{\gamma}+\Gamma_{\gamma \delta}^{\gamma} \Gamma_{\alpha \beta}^{\delta}-\Gamma_{\alpha \delta}^{\gamma} \Gamma_{\gamma \beta}^{\delta}\right)= (\eta'^2-\eta\eta'')\eta^{-4},
\end{equation}

\begin{equation}
    K(\bar{b}) = \det \left(\bar{a}^{-1} \bar{b}\right) =0.
\end{equation}
Thus, we have
\begin{equation}
\label{eq:GMCP}
K({\bar{a}}) - K({\bar{b}}) \neq 0, \quad
    \nabla_{\alpha}\bar{b}_{\beta\gamma}-\nabla_{\beta}\bar{b}_{\alpha\gamma} =0.
\end{equation}
Here, the Christoffel symbol is defined as $\Gamma_{\alpha \beta}^{\gamma} = \frac{1}{2} \bar{a}^{\gamma \delta}(\partial_{\alpha}\bar{a}_{\beta \delta}+\partial_{\beta}\bar{a}_{\alpha \delta}-\partial_{\delta}\bar{a}_{\alpha \beta})$, and $\nabla_{\alpha}(\cdot)$ denotes the covariant derivative. The indices take values in $\{x,~y\}$ and the Einstein summation convention is assumed.

The inequality $K({\bar{a}}) \neq K({\bar{b}})$ violates Gauss's Theorema Egregium, indicating that the reference geometry defined by Eq.~(\ref{eq:refGeo}) is Gauss-incompatible; that is, $\bar{a}$ and $\bar{b}$ together cannot specify a smooth surface. Consequently, minimizing the total strain energy $\mathcal{U}$ under this reference geometry implies that no zero-energy configuration (i.e., $a=\bar{a}, b=\bar{b}$) exists. The actual minimum-energy state must therefore compromise, and cannot satisfy both $a=\bar{a}$ and $b=\bar{b}$ simultaneously—resulting in the typical out-of-plane deformations in a thin, narrow drying leaf (Fig.~\ref{fig:IllustrationOfTheory}). The ensuing shape morphing is governed by the competition between stretching and bending energies, which is moderated by the shrinkage strain $\epsilon_0$, laminae thickness $h$, as well as the midvein radius $R$.

\subsubsection{Planar Configuration} 
\begin{figure}[ht]
    \centering
    \includegraphics[width=.8\textwidth]{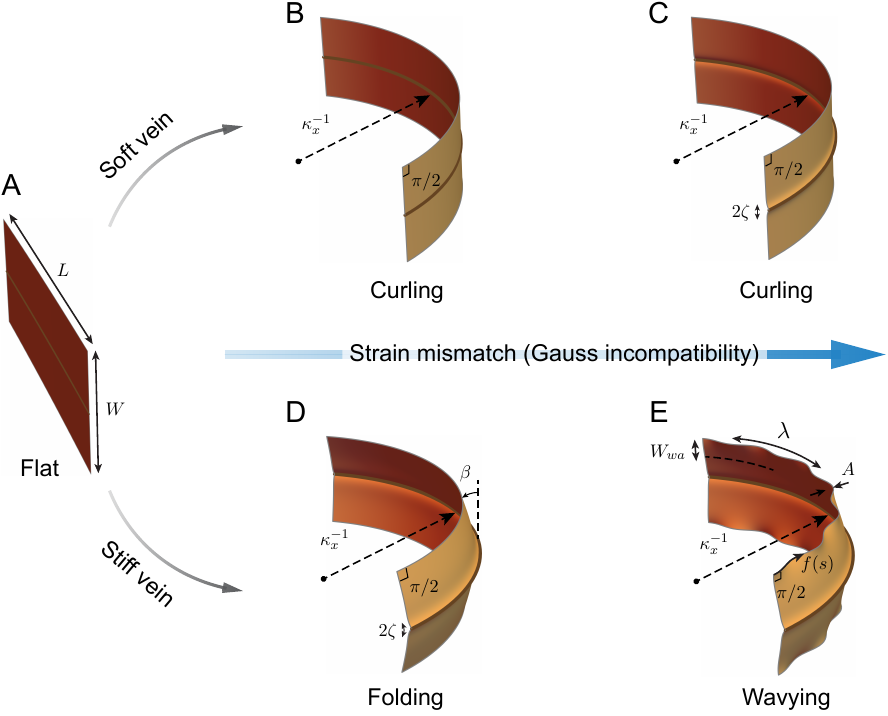}
    \caption{\textbf{Illustration of theoretical predictions of shape formation.} 
    \textbf{A.} Flat configuration. 
    \textbf{B.} Curling configuration with invisible transition layer under small strain mismatch. 
    \textbf{C.} Curling configuration with appreciable transition layer under relatively large strain mismatch.
    \textbf{D.} Folding configuration under small strain mismatch.The small folding angle is denoted as $\beta$.
    \textbf{E.} Waving configuration associated with folding under relatively large strain mismatch. The function of the wavy edges are approximated by $f(s) =A \sin(2\pi/\lambda s)$. Under mild shrinkage strain, the shape-morphing follows the path ``A-B-C" when the midveins are relatively soft, but follows the path ``A-D-E" when the midveins are relatively stiff. In all configurations, we assume that the corner angles remain $\pi/2$ during the drying process.}
    \label{fig:IllustrationOfTheory}
\end{figure}

We begin with the planar configuration (Fig.~\ref{fig:IllustrationOfTheory}A), where the leaf is relatively thick and the strain mismatch is sufficiently small such that the equilibrium state remains nearly flat. In this regime, the second fundamental form vanishes ($b=\bar{b}=0$), and the longitudinal curvature is zero ($\kappa_x = 0$). As a result, the total strain energy is dominated by stretching and scales as

\begin{equation}
    \mathcal{U} = \int_{\mathcal{D}} W_s dS \sim E WLh \epsilon_0^2. 
\end{equation}
This planar configuration is nontrivial, especially when identifying the transition states that precede subsequent shape morphing as the strain mismatch grows.

\subsubsection{Cylindrical Configuration} 
\label{sec:Cconfiguraton}

As the shrinkage strain increases (or equivalently, the leaf thickness decreases), the leaf undergoes buckling proceeding to out-of-plane deformation, eventually adopting a cylindrical configuration, which is referred as \textit{curling-dominated} shape-morphing (see Figs.~\ref{fig:IllustrationOfTheory}B,C). In this surface of revolution, both the first and second fundamental forms become diagonal. Exploiting the mirror symmetry along $y$-direction, the total elastic energy (i.e., Eq. (\ref{eq:totalEnergy0})) can be recast as 

\begin{equation}
    \mathcal{U}=2L \int_{0}^{W/2} \frac{Eh}{8\eta^2}\left((a_{11}-\eta^2)^2+(a_{22}-\eta^2)^2\right)+\frac{Eh^3}{24\eta^2}(b_{11}^2+b_{22}^2)dy.
\end{equation}
To simplify the analysis, we set $\nu = 0$, which does not influence the scaling laws presented in the subsequent sections, and adopt a smooth shrinkage function defined by

\begin{equation}
\label{eq:smoothShrinkage}
    \eta(y) = \frac{1+\eta_0}{2} +\frac{1-\eta_0}{2}\mathrm{tanh}\left(\frac{R-y}{\zeta_0}\right), \quad y \geq 0,
\end{equation}
where $\zeta_0$ characterizes the reference transition width in the $y$-direction, representing a manually set mathematically smooth transition range of shrinkage between the physical midvein and leaf laminae (refer to Fig.~\ref{fig:model}). In the limit $h \to 0$, the actual transition width $\zeta \to \zeta_0$. As described in Section \ref{sec:theory-incompatibility}, the actual configuration's geometry, characterized by $a$ and $b$, must satisfy both Gauss’s Theorema Egregium and the Mainardi–Codazzi equations,

\begin{equation}
K(a) - K(b) = 0, \quad \nabla_{\alpha}b_{\beta\gamma}-\nabla_{\beta}b_{\alpha\gamma} =0.
\end{equation}
By integration, the nonzero (diagonal) components $a_{11}(y)$, $a_{22}(y)$, $b_{11}(y)$ and $b_{22}(y)$ are found to obey
\begin{align}
\label{eq:GMC1}
    b_{11}^2&=ca_{11}-a'_{11}/a_{22},\\
    b^2_{22}&=\frac{1}{b^2_{11}}\left(\frac{a'^2_{11}}{4a_{11}}+\frac{a'_{11}a'_{22}}{4a_{22}}-\frac{a''_{11}}{2}\right)^2 \label{eq:GMC2},
\end{align}
with $c$ being an integral constant~\citep{moshe2013pattern}.

\begin{itemize}
    \item {Small Thickness $h/W\ll1$}.
\end{itemize}
For very thin leaves, the actual metric $a$ is nearly identical to the reference metric $\bar{a}$ over most of the domain, except within the transition layer, where its width $\zeta$ exceeds the reference transition width $\zeta_0$. In principle, the transition layers for $a_{11}$ and $a_{22}$ differ—with $\zeta_{11} \gg \zeta_{22}$. However, following \cite{moshe2013pattern}, we assume that $a_{22}$ is easier to match with $\bar{a}_{22}$, and can be neglected during energy minimization.

We first estimate the stretching energy $\mathcal{U}_S$. Based on Eqs.~(\ref{eq:refGeo}) and~(\ref{eq:smoothShrinkage}), within the transition layer the mismatch in $a_{11}$ between the midvein and the lamina is on the order of  $1-\eta_0$. Consequently, the stretching energy scales as
\begin{equation}
    \mathcal{U}_S = \int_{\mathcal{D}} W_s dS \sim Eh(1-\eta_0)^2\zeta L.
\end{equation}

The bending energy is distributed among the midvein, the lamina, and the transition region. In the lamina region ($y>\zeta$) where $a_{11}$ is nearly constant (that is, $a'_{11} \approx 0$), Eqs.~\ref{eq:GMC1} and~\ref{eq:GMC2} yield 

\begin{equation}
    b^2_{11} \approx c \eta_0^2, \quad b^2_{22} \approx 0
\end{equation}
so that the bending energy in the lamina is
\begin{equation}
    \mathcal{U}_B^{l} = \int_{\mathcal{D}} W_b dS \sim Eh^3cWL.
\end{equation}

Similarly, for the midvein the bending energy is estimated as
\begin{equation}
    \mathcal{U}_B^{m} \sim ER^4c\eta_0^2L.
\end{equation}

Within the transition layer ($0<y<\zeta$), it is straightforward to see that $a_{11} \sim 1$, $a'_{11} \sim (1-\eta_0)/\zeta$ and  $a''_{11} \sim (1-\eta_0)/\zeta^2$. We therefore have
\begin{equation}
\label{eq:bijscaling}
    b^2_{11} \approx c - \frac{(1-\eta_0)^2}{\zeta^2}, \quad b^2_{22} \approx \frac{(1-\eta_0)^2}{b^2_{11}\zeta^4},
\end{equation}
and the corresponding bending energy
\begin{equation}
    \mathcal{U}_B^{t} = \int_{\mathcal{D}} W_b dS \sim Eh^3 \left(c-\frac{(1-\eta_0)^2}{\zeta^2} + \frac{(1-\eta_0)^2}{(c-\frac{(1-\eta_0)^2}{\zeta^2})\zeta^4}\right)\zeta L.
\end{equation}
The total bending energy is then given by
\begin{equation}
    \mathcal{U}_B=\mathcal{U}_B^{l}+\mathcal{U}_B^{m}+\mathcal{U}_B^{t} \sim Eh^3cWL +ER^4c\eta_0^2L+ Eh^3 \left(c-\frac{(1-\eta_0)^2}{\zeta^2} + \frac{(1-\eta_0)^2}{(c-\frac{(1-\eta_0)^2}{\zeta^2})\zeta^4}\right)\zeta L.
\end{equation}
Thus, the total elastic energy is
\begin{equation}
        \mathcal{U}=\mathcal{U}_S+\mathcal{U}_B \sim Eh(1-\eta_0)^2\zeta L + Eh^3cWL + ER^4c\eta_0^2L + Eh^3 \left(c-\frac{(1-\eta_0)^2}{\zeta^2} + \frac{(1-\eta_0)^2}{(c-\frac{(1-\eta_0)^2}{\zeta^2})\zeta^4}\right)\zeta L.
\end{equation}

For a leaf with a narrow midvein ($R\sim h \ll W$) under small shrinkage $(1-\eta_0)$, the transition layer satisfies $\zeta \ll W$. Minimizing the total energy with respect to the integration constant $c$, yields $c= h^2(1-\eta_0)(\sqrt{W\zeta}+W(1-\eta_0))\big/{\zeta^2}$. Substituting this back, the total energy can be expressed as,

\begin{equation}
\label{eq:TotalScaling}
        \mathcal{U} \sim Eh(1-\eta_0)^2\zeta L + \frac{Eh^3L(1-\eta_0)W^{1/2}}{\zeta^{3/2}}\left(1+\sqrt{\frac{(1-\eta_0)^2 W}{\zeta}}\right).
\end{equation}
Here, the first term corresponds to the stretching energy scaling, while the second term represents the bending energy scaling. The width of transition layer can be determined by balancing $\mathcal{U}_{B}$ and $\mathcal{U}_{S}$: $\zeta \sim W^{1/3}h^{2/3}$ for $\zeta/W \ll (1-\eta_0)^2$; $\zeta \sim (1-\eta_0)^{-2/5}W^{1/5}h^{4/5}$ for $\zeta/W \gg (1-\eta_0)^2$. Accordingly, the energy scaling in Eq.~(\ref{eq:TotalScaling}) becomes

\begin{numcases}{\mathcal{U}\sim}
EL\epsilon_0^2 W^{1/3} h^{5/3} \quad \text {for} ~\zeta/W \ll \epsilon_0^2, \label{eq:Total1} \\
EL\epsilon_0^{8/5} W^{1/5} h^{9/5} \quad \text {for} ~\zeta/W \gg \epsilon_0^2. \label{eq:Total2}
\end{numcases}
and the corresponding curvature of midvein (refer to Eq.~(\ref{eq:bijscaling})) scales as

\begin{numcases}{\kappa_x \sim b_{11} \sim}
\epsilon_0 W^{-1/3} h^{-2/3} \quad \text {for} ~\zeta/W \ll \epsilon_0^2, \label{eq:kxSmallh} \\
\epsilon_0^{4/5} W^{-2/5} h^{-3/5} \quad \text {for} ~\zeta/W \gg \epsilon_0^2. \label{eq:kxLargeh}
\end{numcases}
Here, $\epsilon_0 = 1-\eta_0$ denotes the applied thermal strain (shrinking strain). These two expressions delineate distinct mathematical regimes. However, observations of natural leaves and the geometric consideration of our model show that the scaling in Eq.~(\ref{eq:kxSmallh}) is the most pertinent. A detailed discussion is provided in Section \ref{sec:validationscaling}.

\begin{itemize}
    \item {Vanishing Thickness $h/W \to 0$}.
\end{itemize}
For an infinitely thin leaf, an isometric embedding of the reference metric $\bar{a}$ minimizes the total energy. In this limit, $a=\bar{a}$ and the stretching energy $W_s \sim 0$. By rescaling Eqs.~(\ref{eq:refGeo}), (\ref{eq:GMC1}), and (\ref{eq:GMC2}), the second fundamental form is obtained as

\begin{equation}
\label{eq:bSoR}
    b_{11} = \sqrt{c\eta^2-\eta'^2}, \quad b_{22} = \frac{\eta'^2-\eta \eta''}{\sqrt{c\eta^2-\eta'^2}}.
\end{equation}

Substituting Eq.~(\ref{eq:smoothShrinkage}) into Eq.~(\ref{eq:bSoR}) and evaluating at $y=R$, provides a lower bound for $c$: $c \geq \frac{(1-\eta_0)^2}{\zeta_0^2(1+\eta_0)^2}$. For sufficiently small $\zeta_0$ and shrinkage, the curvature in the bulk region scales as
\begin{equation}
\label{eq:kxZeroh}
    \kappa_x \sim b_{11} \sim \sqrt{c \eta_0^2} \sim \zeta_0^{-1}\epsilon_0.
\end{equation}
and the bending energy scales as
\begin{equation}
    \mathcal{U} =  \int_{\mathcal{D}} W_b dS \sim EWLh^3 \zeta_0^{-2}\epsilon_0^2. 
\end{equation}

From the combined analysis of Eqs.~(\ref{eq:kxSmallh}) and (\ref{eq:kxZeroh}), and under the assumptions of narrow transition layers, we conclude that, in both the small- and vanishing-thickness limits, the longitudinal curvature in the buckled configuration scales linearly with the shrinkage strain, i.e.,
\begin{equation}
\label{eq:CurvatureScaling}
    \kappa_x \sim \epsilon_0.
\end{equation}

For a drying leaf with finite thickness, there exists a critical shrinkage strain. At this critical point, the initially planar configuration transforms into a surface of revolution with a relatively broad transition layer (see Fig.~\ref{fig:IllustrationOfTheory}C). Balancing the competing energies yields the critical strain
\begin{equation}
\label{eq:CritStrainScaling}
    \epsilon_0^c \sim h^2W^{-2},
\end{equation}
which aligns with the geometric intuition of Gauss incompatibility—that is, as the leaf becomes infinitely thin ($h \to0$), the critical strain $\epsilon_0^c$ approaches zero. Both the scaling of midvein curvature and critical strain will be validated in Section \ref{sec:validationscaling}.

\subsubsection{Folding and Wavy Configuration}  \label{sec:wavy}
The intrinsic geometric incompatibility in the leaf persists throughout drying, but the way it is accommodated evolves with increasing shrinkage. Our non‑Euclidean elasticity framework predicts that, as \(\epsilon_0\) grows, new morphologies emerge from a combination of geometric and energetic trade‑offs. We therefore adopt a geometrico‑mechanical perspective under the assumption of small relative thickness \(h/W\). Assuming that the corner turning angles of the rectangular domain \(\mathcal{D}\) remain \(\pi/2\) during shrinkage (as shown in Fig.~\ref{fig:IllustrationOfTheory}B-E), the Gauss–Bonnet theorem dictates a global balance between the integrated Gaussian curvature and the boundary’s geodesic curvature~\citep{do2016differential}, i.e.,

\begin{equation}
\label{eq:GB}
   \int_{\partial \mathcal{D}} k_g dl \sim - \int_{\mathcal{D}} K dS 
\end{equation}

At low shrinkage, the overall geometric incompatibility is relatively mild such that the accumulated Gaussian curvature near the midvein is small.  Regions away from the midvein (including the free edges) can then be nearly isometrically embedded, yielding an almost cylindrical shape with negligible boundary geodesic curvature, $
k_g \sim \frac{1}{L}\int_{\mathcal{D}}K\,dS \ll 1$, 
consistent with the cylindrical configuration of Section~\ref{sec:Cconfiguraton}.
\begin{itemize}
    \item {Mechanism of Folding and Wavy}.
\end{itemize}

As shrinkage increases, geometric incompatibility near the midvein becomes more pronounced. At the same time, the midvein region accumulates significant Gaussian curvature, as indicated by the growth of \(\int_{\mathcal{D}} K\, dS\). According to Eq.~(\ref{eq:GB}), this must be compensated by an increase in the boundary geodesic curvature, \(\int_{\partial \mathcal{D}} k_g\, dl\). Geometrically, the most direct way to achieve this is for the lamina to develop a fold perpendicular to the midvein, forming a frustum-like profile (Fig.~\ref{fig:IllustrationOfTheory}D). This localized bending increases the geodesic curvature along the edge, partially resolving the underlying metric incompatibility.
Mechanically, introducing such a fold converts energetically costly stretching into relatively inexpensive bending. For \(h/W \ll 1\), stretching energy (which scales with \(h\)) is far more expensive than bending energy (which scales with \(h^3\)). In the vicinity of a stiff midvein, the rapid growth of Gaussian curvature leads to a corresponding accumulation of stretching energy, \(\mathcal{U}_s \sim \int_{\mathcal{D}}(\Delta^{-1}K)^2\, dS \sim \int_{\mathcal{D}}(\zeta_0^2 K)^2\, dS\). To reduce its total energy, the system therefore prefers to adopt a fold perpendicular to the midvein, as shown in Fig.~\ref{fig:IllustrationOfTheory}D.

With further shrinkage, the fold angle \(\beta\) increases, while the effective length of each parallel remains approximately fixed (\(\sim L\eta_0\)). This amplifies the uncompensated compressive strain along the longitudinal free edge, given by \(\varepsilon_{\rm un} \sim \kappa_x W \beta\). Relying solely on the folding mechanism to accommodate the integrated Gaussian curvature would require large local compressive strains along the edge, thereby incurring a substantial stretching energy penalty—especially for thin leaves. Instead, the system further alleviates the mismatch by developing a wavy pattern along the leaf edge, introducing more degrees of freedom (Fig.~\ref{fig:IllustrationOfTheory}E). This wavy configuration both enhances the local geodesic curvature (to satisfy Gauss–Bonnet) and confines bending to the edge region, thereby further reducing in-plane stretching. The formation and evolution of both the folding and wavy configurations are referred as \textit{folding-dominated} shape-morphing.

Note that the shape morphing of drying leaves fundamentally originates from inherent Gaussian incompatibility (see Section~\ref{sec:theory-incompatibility}). As a result, the same sequence of morphological transitions can be reproduced—even under a fixed shrinkage ratio \((1 - \eta_0)\)—simply by increasing the bending stiffness ratio of the midvein to the lamina. This demonstrates that the morphing pathway is governed not only by shrinkage magnitude but also by stiffness contrast. Such sensitivity offers a route to programmability through internal structural design.

\begin{itemize}
    \item {Scaling Argument for Wavy Configurations}.
\end{itemize}

To gain insight into the wavy‐edge morphology, we present a rough scaling argument in the thin‐plate limit $h/W \ll 1$ and for small fold angles $\beta \ll 1$, where the fold regime is vanishingly narrow and the entire sheet—including the wavy zone—remains nearly isometric.

Let the free edge follow a sinusoidal profile $f(s) \sim A \sin(2\pi/\lambda s)$ over a transverse width $W_{wa}$, where $s$ runs along the midvein, $A$ is the wrinkle amplitude, and $\lambda$ its wavelength. In the wavy zone, the dominant curvature scales as $\kappa_{wa} \sim A/\lambda^2$, so its bending energy satisfies $\mathcal{U}_{wa} \sim E h^3  \int \kappa_{wa}^2 dS \sim E h^3A^2 \lambda^{-4} L W_{wa}$. Meanwhile, approximate isometry demands that the local slope satisfies $ A^2/\lambda^2 \sim \epsilon_0$. Since $\beta \ll 1$, the remainder of the sheet—including the inconsiderably folded or cylindrical part—continues to carry both stretching and bending energy $\mathcal{U}_{cy} \sim  EL\epsilon_0^2 (W-W_{wa})^{1/3} h^{5/3}$ in agreement with Eq.~(\ref{eq:Total1}). For the wavy pattern to be energetically favourable and stable, one expect that $\mathcal{U}_{wa} \sim \mathcal{U}_{cy}$ with $W_{wa} \sim W$. The above arguments yield a rough estimation of the characteristic wavelength of the wavy pattern, $\lambda \sim h^{2/3}\epsilon_0^{-1/2} W^{1/3}$.

This scaling suggests that the wavy wavelength grows with the lamina thickness $h$ and leaf width $W$, but decreases as the applied strain $\epsilon_0$ increases. It also indicates that, for sufficiently large $h$ or small $L$ (so that $\lambda > L$), stable wavy patterns become difficult to observe.

\subsubsection{Validation of scaling of midvein curvature and critical strain}
\label{sec:validationscaling}

\begin{figure}[ht]
    \centering
    \includegraphics[width=1.0\textwidth]{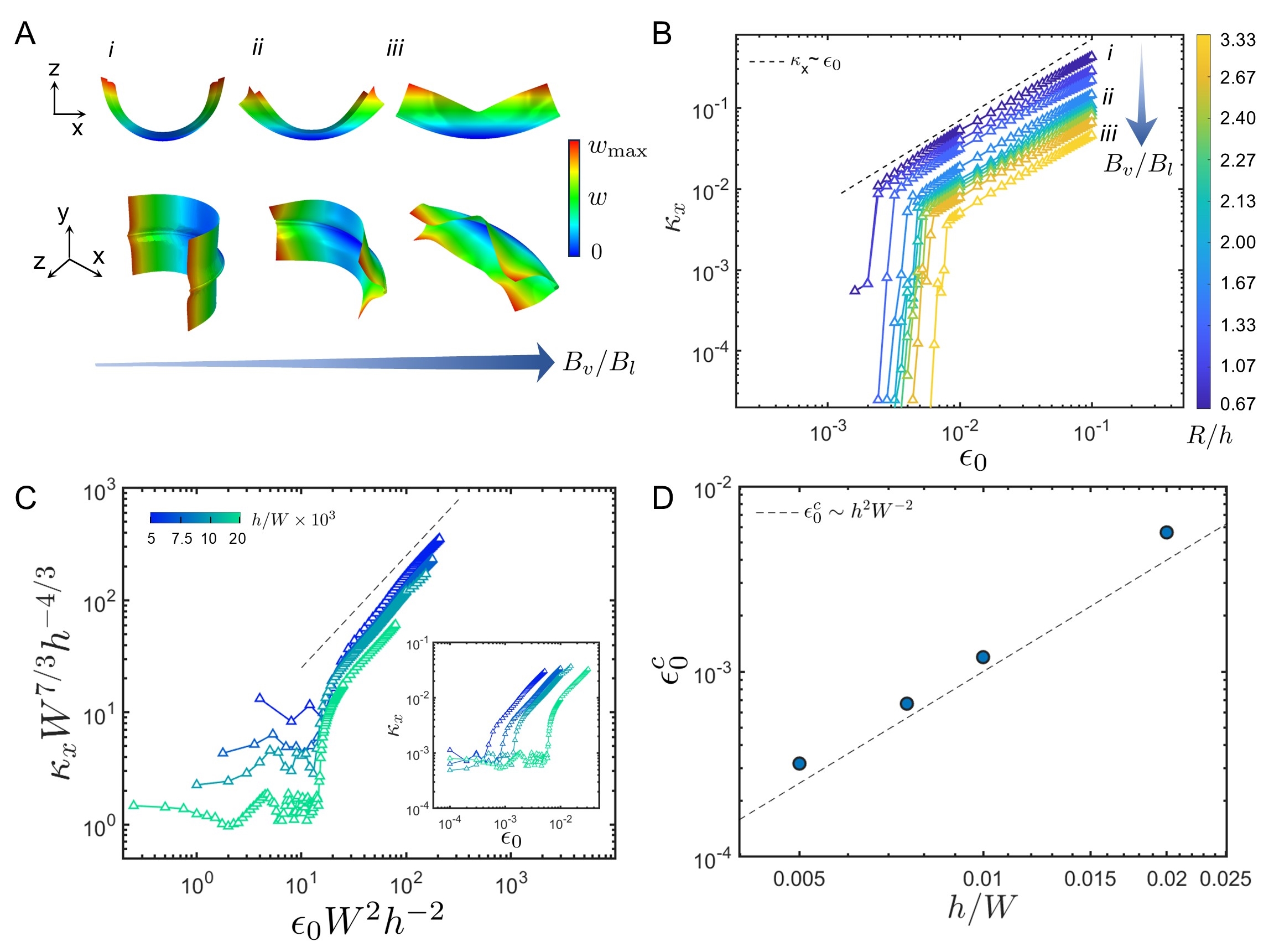}
\caption{\textbf{Scaling of midvein curvature and critical strain.}  
\textbf{A.} Simulations of leaves with fixed geometry (\(L/W = 2\), \(h/W = 7.5\times10^{-3}\)) and varying midvein radius \(R/h =\) (i) 0.10, (ii) 0.16, (iii) 0.25. The arrow indicates increasing midvein bending stiffness \(B_v/B_l\). Top and bottom rows show two different views, as indicated by the coordinate axes. Contours represent the out-of-plane displacement \(w\) at \(\epsilon_0 = 0.1\).  
\textbf{B.} Midvein curvature \(\kappa_x\) as a function of applied shrinkage strain \(\epsilon_0\). Colors denote different \(R/h\) ratios corresponding to increasing \(B_v/B_l\), with (i–iii) matching the configurations shown in \textbf{A}. The dashed line represents the scaling relation predicted by Eq.~(25).  
\textbf{C.} Rescaled midvein curvature \(\kappa_x W^{7/3} h^{-4/3}\) plotted against rescaled shrinkage strain \(\epsilon_0 W^2 h^{-2}\). The inset shows the corresponding unscaled data. All simulations maintain a constant \(R/h = 0.5\) while varying \(h/W\).  
\textbf{D.} Critical shrinkage strain \(\epsilon_0^c\) extracted from \textbf{C} as a function of the thickness-to-width ratio \(h/W\). Dashed lines in \textbf{C} and \textbf{D} represent theoretical scaling predictions.}
    \label{fig:curvature}
\end{figure}

To investigate how midvein stiffness influences shape formation, we performed FEM simulations on leaves with varying midvein radius, quantified by $R/h$ and corresponding to $B_v/B_l$ values ranging from 0.3 to 100. Figure~\ref{fig:curvature}A shows three representative deformed configurations (i–iii), illustrating that increasing the midvein radius suppresses curvature, as a stiffer midvein more effectively resists bending. This trend is quantified in Fig.~\ref{fig:curvature}B, where the midvein curvature $\kappa_x$ is plotted against the applied shrinkage strain $\epsilon_0$. At a fixed strain, $\kappa_x$ decreases with increasing $B_v/B_l$. More importantly, despite differences in midvein rigidity, all curves exhibit a consistent logarithmic slope of 1 in the post-buckling regime, indicating that curvature scales linearly with strain (i.e., $\kappa_x \sim \epsilon_0$). This scaling behavior holds robustly across all geometries studied here and agrees with theoretical predictions from Eq.~(\ref{eq:CurvatureScaling}), confirming its validity in the thin-sheet regime with narrow transition zones.

We further examined the effect of lamina thickness by fixing the midvein geometry ($R/h = 0.5$) and varying the thickness-to-width ratio $h/W$. As shown in Fig.~\ref{fig:curvature}C, increasing the lamina thickness reduces curvature. When rescaled according to Eq.~(\ref{eq:CurvatureScaling}), the curvature data collapse onto a single curve, reinforcing the universality of the scaling law. From this master curve, we extracted the critical shrinkage strain $\epsilon_0^c$ corresponding to the onset of buckling. Figure~\ref{fig:curvature}D shows that $\epsilon_0^c$ scales quadratically with $h/W$ (i.e., $\epsilon_0^c \sim h^2/W^2$), in agreement with Eq.~(\ref{eq:CritStrainScaling}). Together, these results establish a coherent framework linking curvature generation and instability onset to geometric and material parameters, offering predictive insights into the morphing mechanisms of midvein-constrained leaves, and suggesting broader principles for the mechanical design of adaptive thin structures.

Note that the deformation of the leaf is also influenced by the midvein radius (Fig.~\ref{fig:curvature}A). As midvein rigidity increases, the midvein develops smaller curvature, while the lamina becomes more folded, accompanied by amplified edge waviness. As discussed in Section~\ref{sec:wavy}, the edge waviness arises from the need to compensate for accumulated Gaussian curvature, as described by the Gauss-Bonnet theorem, reducing stretching energy by converting it into localized bending along the edge and thereby alleviating geometric incompatibility. The wavelength of the waviness increases with the thickness and width of the lamina, but decreases with the shrinkage strain. Figure~\ref{fig:evolve} illustrates the morphological differences between leaves without and with edge waviness.

\subsection{Phase diagram of morphing modes}

\begin{figure}[ht]
    \centering
    \includegraphics[width=0.6\textwidth]{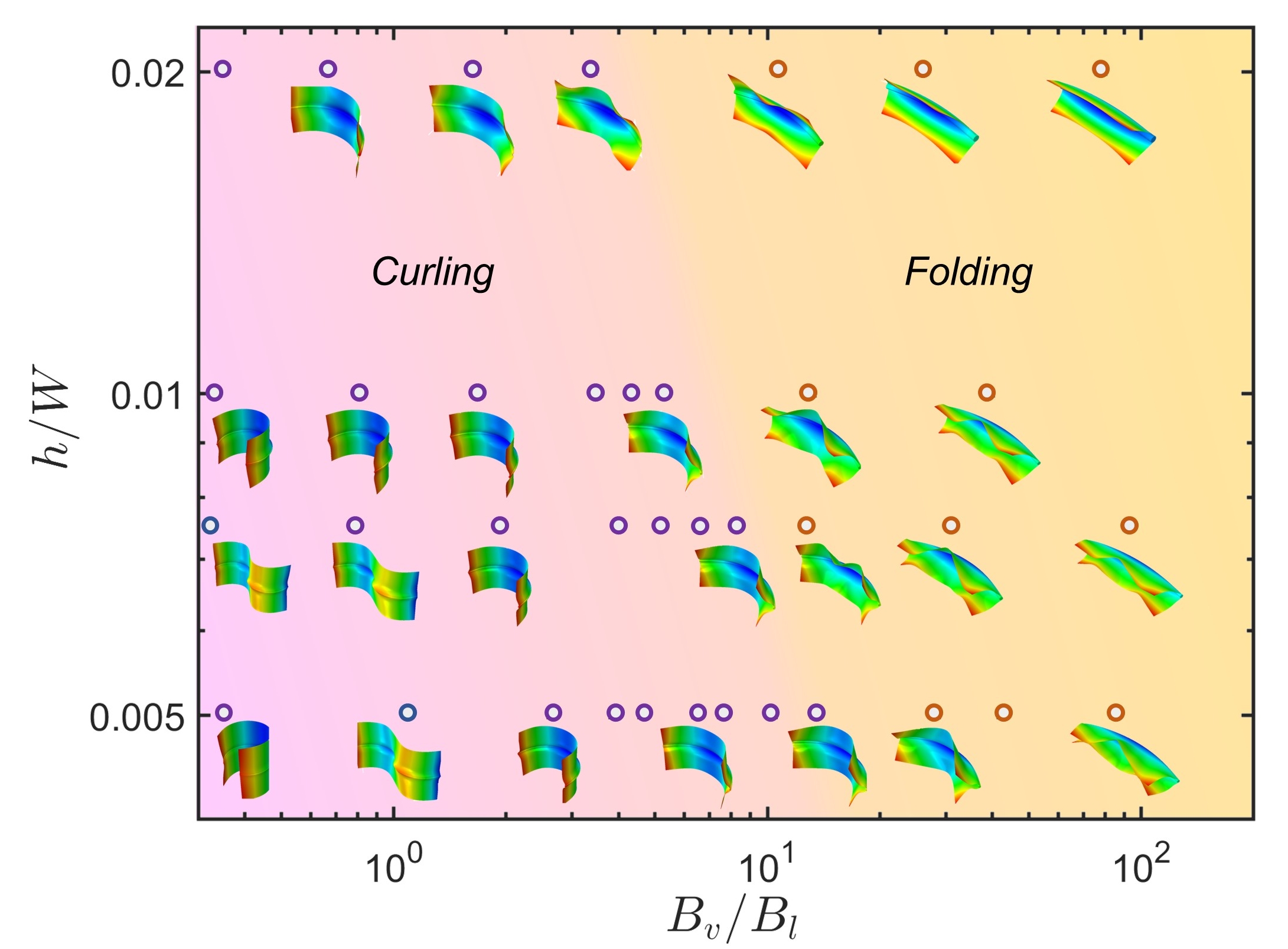} 
    \caption{\textbf{Phase diagram of the shape modes.} 
    The shape modes change with bending stiffness ratio of the midvein and the lamina, $B_v/B_l$. The thickness-to-width ratio of the lamina, $h/W$, change the edge waviness of the lamina after folding. The red and yellow regions indicate the curling-dominated and folding-dominated regimes, respectively. Note that the transition is a gradual process. }
    \label{fig:phase}
\end{figure}

We have shown that two distinct modes of equilibrium configurations can emerge depending on the relative rigidity of the two components (Fig.~\ref{fig:evolve} and Fig.~\ref{fig:curvature}A): the \textit{curling-dominated} mode characterized by a cylindrical configuration with curvature along the longitudinal direction; and the \textit{folding-dominated} mode marked by prominent transverse folding of the lamina about the midvein. Our FEM simulations reveal that the bending rigidity ratio between the midvein and the lamina, $B_v/B_l$, is a key regulatory parameter governing the formation and evolution of these configurations. Figure~\ref{fig:phase} presents a phase diagram in the space of $B_v/B_l$ and $h/W$, delineating the regimes of different morphologies. In general, as the midvein becomes  stiffer (from left to right in the diagram), the system compensates by reducing midvein curvature and transitioning from a \textit{curling-dominated} to a \textit{folding-dominated} configuration. The emergent morphologies and their transitions are also significantly influenced by the normalized lamina thickness, $h/W$. As $h/W$ increases (from bottom to top), the longitudinal curvature in the curling-dominated state steadily decreases, and the transition boundary—quantified by $B_v/B_l$—shifts to lower values. This shift arises because longitudinal curling becomes energetically more costly in thicker leaves, whereas transverse folding remains comparatively favorable.

It is worth noting that when the rigidity ratio is lower than one—i.e., the midvein is less rigid than or comparable to the lamina—the system tends to \textit{randomly} exhibit a higher-order \textit{curling} mode, namely an \textit{S}-curled shape (bottom-left region of the phase diagram in Fig.~\ref{fig:phase}). This occurs because the strain energies of the \textit{S}-curled and \textit{C}-curled modes are nearly identical under these conditions, a phenomenon that will be further discussed in Section~\ref{sec:discussion}. As the rigidity ratio increases, the energy difference between the two modes becomes increasingly pronounced; however, it nearly vanishes when the midvein diameter approaches the lamina thickness. This trend highlights the critical role of mechanical contrast between the midvein and lamina in determining the final morphology.
According to the theoretical analysis in Section~\ref{sec:wavy}, when the midvein is sufficiently stiff and the shrinkage strain is large—leading to higher Gaussian curvature incompatibility—edge waviness emerges as a secondary instability. This wavy configuration enhances the boundary geodesic curvature to compensate for the increasing Gaussian curvature near the midvein, thereby mitigating geometric frustration in the folding-dominated regime. Moreover, our theory predicts that the wavelength of the waviness increases with the normalized leaf thickness, consistent with the phase diagram: stable wavy patterns are more readily observed in thinner leaves, whereas they are suppressed in thicker ones under otherwise identical geometric conditions.
Taken together, these findings establish a predictive framework linking mechanical heterogeneity to morphological diversity in natural thin structures, providing new insights into the physical principles underlying shape formation in biological systems.

\section{Discussion}\label{sec:discussion}
Previous research has primarily focused on shape morphogenesis driven by mechanical buckling of the leaf blade due to differential growth or expansion (i.e., positive strain) \citep{sharon2007,maha2009-longleaf,huang2018-differential,xu2020-lotus}. In contrast, the present study investigates the role of differential shrinkage (negative strain) induced by drying-related tissue contraction as a mechanism for generating curved leaf geometries. Despite differing in the sign of the strain, both expansion- and contraction-driven morphogenesis fundamentally rely on in-plane strain mismatches to induce buckling behavior \citep{guo2025localized}, producing curvature regardless of strain polarity.
The resulting morphologies, however, differ markedly between the two regimes. In growth-driven systems, higher strain at the leaf margins (i.e., a positive strain gradient from midvein to edge) typically leads to twisting or saddle-like geometries \citep{maha2009-longleaf, huang2018-differential, xu2020-lotus}, while a negative strain gradient—where the margins expand less than the center—often gives rise to center-bulged configurations. These distinct configurations arising from varying strain gradients have been qualitatively mapped in a phase diagram (see Fig.~8 in \citep{guo2022}).
Here, we identify a representative biological system—drying leaves—that morphs under a negative strain gradient from midvein to edge via differential shrinkage. We demonstrate that such shrinkage-induced strains result in global curvature along the midvein and folding of the lamina, distinct from the morphologies produced by edge-dominated growth. These findings offer a complementary perspective to classical growth-based models of leaf morphogenesis.

\begin{figure}[h]
    \centering
    \includegraphics[width=0.6\textwidth]{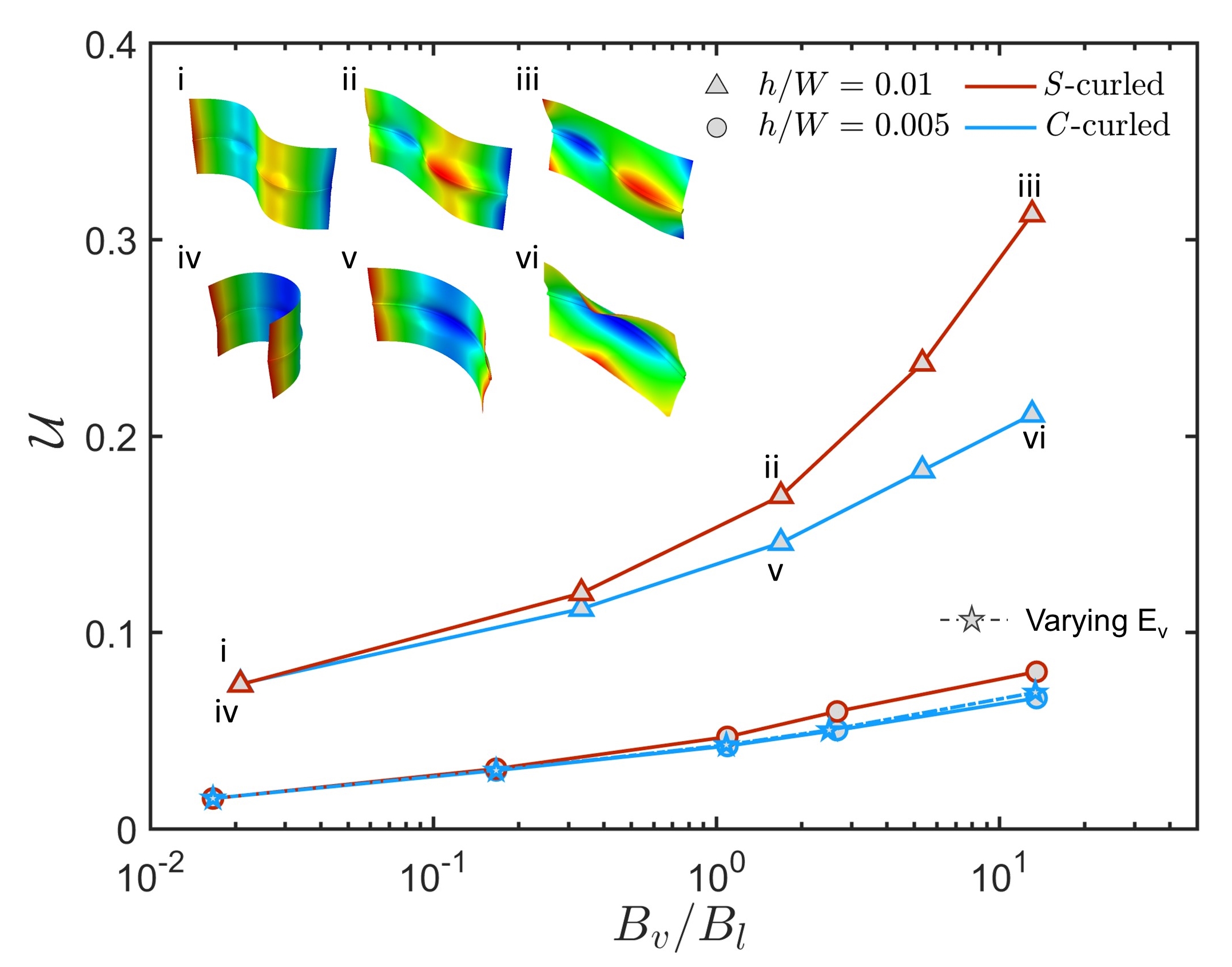} 
    \caption{\textbf{Comparison of elastic energy between \textit{S}- and \textit{C}-curled modes.} 
    Each mode of the geometric imperfections is applied to the initial configurations, and thus the corresponding buckling modes emerged in the simulation of shrinking process. Data is shown for elastic strain energy at $\epsilon_0=0.05$. Dotted curves, $L/W=2,h/W=0.01$. Solid curves, $L/W=2,h/W=0.005$. For the curve with marker symbol ``$\star$'', the Young's modulus of the vein $E_v$ is varied, while for the other cases $R$ is varied.}
    \label{fig:energy}
\end{figure}

Another notable observation in this study is the emergence of a higher-order mode—the \textit{S}-curled shape—within the \textit{curling}-dominated regime, alongside the more common \textit{C}-curled pattern. This phenomenon is evident in both real leaves (Fig.~\ref{fig:overview}C) and simulations (Fig.~\ref{fig:overview}D and Fig.~\ref{fig:phase}). Figure~\ref{fig:energy} compares the elastic strain energies of the \textit{S}- and \textit{C}-curled configurations, demonstrating how the rigidity ratio governs the energetic preference between the two modes.
As the rigidity ratio increases—either via a stiffer midvein modulus ($E_v$) or a larger midvein radius ($R$)—the energy contrast between the two morphologies becomes more pronounced, favoring the \textit{C}-mode. Conversely, when the midvein's rigidity approaches or falls below that of the lamina, the energy difference diminishes, leading to random mode selection driven by imperfections or external perturbations.
This behavior can be explained by the beam’s bending resistance: when the midvein is compliant, it offers little resistance, and both \textit{C}- and \textit{S}-modes have comparable elastic energies. In contrast, a stiff midvein penalizes the \textit{S}-mode due to the high curvature required in opposing directions, which increases its bending energy. The \textit{C}-mode, by involving a more uniform curvature, incurs a lower energy cost for the same beam stiffness, while the flexible lamina accommodates both modes with minimal energy penalty.
Our data also highlight the role of geometry: thinner laminae ($h/W = 0.005$) exhibit smaller energy differences between the modes compared to thicker ones ($h/W = 0.01$), underscoring the coupled influence of material and geometric parameters in determining buckling pathways.

In this study, the proposed mechanism underscores a universal principle for driving leaf morphing that does not rely on detailed cellular structures, provided a midvein is present. Nevertheless, certain cellular features in real leaves can influence the preferential selection of morphing modes. For example, the midrib often protrudes from the lamina plane, introducing asymmetry in the thickness direction \citep{araujo2014leaf}. This asymmetry may promote bending toward either the adaxial or abaxial side, resulting in \textit{C}- or \textit{S}-shaped configurations. In-plane asymmetries in midvein thickness may also facilitate folding. Additionally, shrinkage of collenchyma cells surrounding the vein can induce folding behavior, functioning analogously to a mechanical hinge \citep{guo2024dehydration}.
It is also important to note that not all leaves possess prominent midveins—for instance, ginkgo and some monocotyledonous angiosperms such as palms. The post-drying deformation of such leaves differs markedly from the midvein-governed cases discussed here. Instead of midvein-induced curving, they often exhibit rolling \citep{kadioglu2012} or hinge-like folding \citep{guo2024dehydration}, driven by specific cellular architectures. These examples highlight the critical role of venation—particularly the midvein—in regulating leaf morphologies.

Although the morphological transformation of drying leaves is not an active biological process, it contributes significantly to ecosystem functions such as nutrient cycling, habitat formation, and environmental protection \citep{biviano2025settling}. During senescence—a genetically regulated phase of aging and nutrient remobilization—leaves shrink and deform into irregular structures that help retain moisture, trap soil particles, and foster microbial activity. These physical deformations accelerate decomposition, enriching the soil with organic matter. Moreover, the resulting forms provide microhabitats for small organisms such as insects and fungi, thereby enhancing biodiversity, and may offer protection to seeds or spores by buffering against wind and rain.

\section{Conclusion}\label{sec:conclusion}
\begin{figure}[h]
    \centering
    \includegraphics[width=\linewidth]{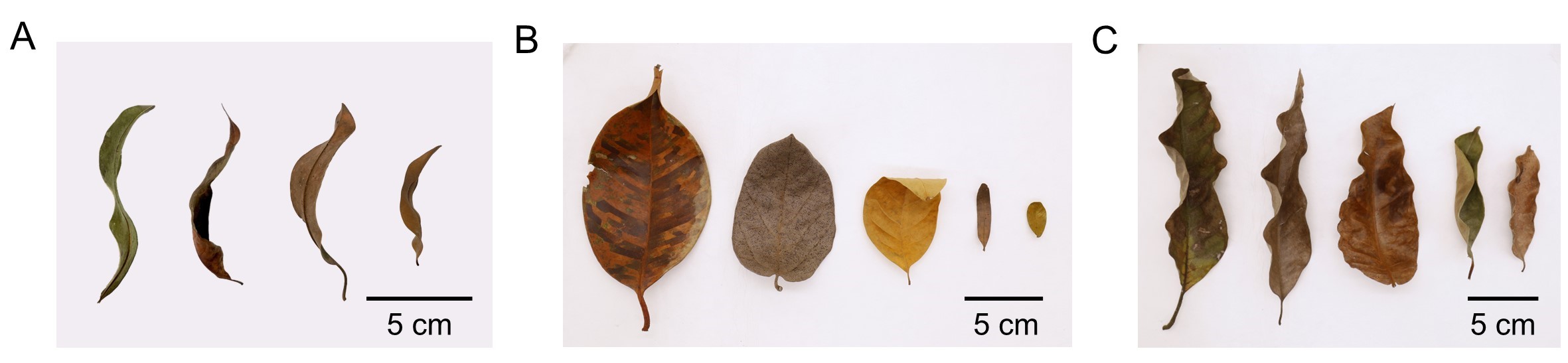}
    \caption{\textbf{Typical configurations observed in natural drying leaves, corresponding to the morphologies discussed in this study.} 
    \textbf{A.} \textit{Curling-dominated} configuration with an \textit{S}-curled shape. \textbf{B.} \textit{Curling-dominated} configuration with a \textit{C}-curled shape. \textbf{C.} \textit{Folding-dominated} configuration characterized by transverse folds and edge waviness.}
    \label{fig:si-leaf}
\end{figure}

In this study, we demonstrate that the diverse and intriguing morphologies of drying leaves arise from the mechanical constraints imposed by venation on lamina contraction during senescence or dehydration. As the leaf dries, the lamina undergoes shrinkage, while the midvein (typically stiffer) resists deformation, leading to strain mismatch in leaf tissues. The resulting morphologies are governed by the interplay between material properties and geometric parameters---our analysis and simulations reveal two distinctive types of configurations that emerge depending on the relative bending rigidity between the lamina and the midvein: \textit{curling-dominated} and \textit{folding-dominated} morphologies. In the curling-dominated regime, the leaf exhibits either a \textit{C}-curled or a higher-order \textit{S}-curled shape, both commonly observed in nature. As the rigidity contrast increases (i.e., as the midvein becomes stiffer relative to the lamina), the system transitions to a folding-dominated morphology.
To validate these findings, we performed simulations using plate models of elliptical laminae coupled to midveins, where the midvein is represented as a stiffened region with increased thickness and elastic modulus. Remarkably, drying leaves collected from various species show strikingly similar morphologies. As shown in Fig.~\ref{fig:si-leaf}, most leaves can be categorized into these two types: curling-dominated configurations with either \textit{S}-curled (Fig.~\ref{fig:si-leaf}A) or \textit{C}-curled (Fig.~\ref{fig:si-leaf}B) shapes, and folding-dominated configurations characterized by transverse folds and edge waviness (Fig.~\ref{fig:si-leaf}C). However, due to the relatively small thickness of real leaves, pure folded configurations are rarely observed; instead, folding is typically accompanied by edge waviness, which helps relieve geometric frustration.
Additionally, comparison between Fig.~\ref{fig:si-leaf}A and Fig.~\ref{fig:si-leaf}B suggests that leaves with larger length-to-width ratios are more likely to develop \textit{S}-curled shapes, indicating the geometric aspect ratio as another important factor influencing morphogenesis.
These observations reinforce the relevance of our mechanical framework and provide new insights into how the internal structure of leaf blades dictates morphological transformations during drying.

This study provides a general framework for understanding the emergence of representative drying leaf morphologies; however, several structural and environmental factors not considered here may further influence shape formation. For example, midvein tapering—where the midvein narrows along its length—could affect curvature development during drying. Additionally, larger leaves often possess thicker midveins, potentially impacting buckling behavior and critical strain thresholds. 
Furthermore, the interactions between the midvein and the lamina are complex. The compressive loads that induce the buckling of the midvein are generally along the longitudinal direction as the midvein deforms, i.e., they are \textit{follower loads} \citep{Bolotin1963}. How a beam subjected to follower loads buckles is a topic requiring further research. Variations in cellulose, lignin, and water content across plant species further contribute to differences in deformation behavior, alongside external factors such as wind exposure and surface contact. Future work could systematically investigate these factors to refine our understanding of how venation patterns and material composition govern drying leaf morphologies.
Compared to previous studies, our findings provide new insights into the mechanics underlying leaf shape transformations, bridging the gap between biological structure and physical deformation. This mechanistic understanding not only advances fundamental knowledge in plant morphogenesis but also offers inspiration for the design of bio-inspired materials, self-shaping structures, and programmable soft matter systems \citep{box2022gigantic,huang2024integration,guo2025localized}.

\appendix

\section*{CRediT authorship contribution statement}

\textbf{Kexin Guo}: Writing – review \& editing, Writing – original draft, Visualization, Validation, Software, Methodology, Investigation, Formal analysis, Data curation, Conceptualization. \textbf{Yafei Zhang}: Writing – review \& editing, Writing – original draft, Visualization, Validation, Methodology, Investigation, Formal analysis, Data curation. \textbf{Massimo Paradiso}: Writing – review \& editing, Validation, Formal analysis. \textbf{Yuchen Long}: Writing – review \& editing, Validation, Formal analysis. \textbf{K. Jimmy Hsia}: Writing – review \& editing, Investigation, Supervision, Project administration, Funding acquisition, Conceptualization. \textbf{Mingchao Liu}: Writing – review \& editing, Methodology,  Investigation, Supervision, Project administration, Funding acquisition, Conceptualization.

\section*{Declaration of competing interest}

The authors declare that they have no known competing financial interests or personal relationships that could have appeared to influence the work reported in this paper.

\section*{Acknowledgments}

K.G., M.P. and K.J.H. acknowledge the financial support of the Ministry of Education, Singapore, via MOE AcRF Tier 3 Award MOE-MOET32022-0002. Y.L acknowledges the start-up funding from the National University of Singapore, under the Prestige Young Professorship programme. M.L. acknowledges the start-up funding from the University of Birmingham. We are grateful for helpful discussions with Cyrus Mostajeran of Nanyang Technological University and Dominic Vella of University of Oxford.

\section*{Data and code availability}

The datasets and code generated during the current study are available from the corresponding author upon reasonable request.

\section*{Declaration of generative AI and AI-assisted technologies in the writing process}
During the preparation of this work, the authors used OpenAI’s ChatGPT-4o to improve the language and readability of the text. After using this tool, the authors reviewed and edited the content as needed and take full responsibility for its publication.


\end{document}